\documentstyle{mn}

\input{epsf}

\input{epsf}

\begin{document}

\title[Cosmological Perturbation Theory and the Spherical Collapse model]
{Cosmological Perturbation Theory and the Spherical Collapse model-
II. Non-Gaussian initial conditions.} 

\author[E. Gazta\~{n}aga and P. Fosalba]
{Enrique Gazta\~{n}aga and Pablo Fosalba \\ 
Institut d'Estudis Espacials de Catalunya, Research Unit (CSIC), \\
Edf. Nexus-201 - c/ Gran Capit\`a 2-4, 08034 Barcelona, Spain}

\maketitle
 
\def\Mpc{{\,h^{-1}\,{\rm Mpc}}}
\def\mpc {h^{-1} {\rm{Mpc}}}
\def\and  {\it {et al.} \rm}
\def\rmd {\rm d}

\def\etal{{\it et al. }}
\def\ie {{\it i.e., }} 
\def\eg {{\it e.g., }}
\def\spose#1{\hbox to 0pt{#1\hss}}
\def\simlt{\mathrel{\spose{\lower 3pt\hbox{$\mathchar"218$}}
     \raise 2.0pt\hbox{$\mathchar"13C$}}}
\def\simgt{\mathrel{\spose{\lower 3pt\hbox{$\mathchar"218$}}
     \raise 2.0pt\hbox{$\mathchar"13E$}}}
\def\beq{\begin{equation}}
\def\eeq{\end{equation}}
\def\bce{\begin{center}}
\def\ece{\end{center}}
\def\bea{\begin{eqnarray}}
\def\eea{\end{eqnarray}}
\def\ben{\begin{enumerate}}
\def\een{\end{enumerate}}
\def\ul{\underline}
\def\ni{\noindent}
\def\nn{\nonumber}
\def\bs{\bigskip}
\def\ms{\medskip}
\def\wt{\widetilde}
\def\brr{\begin{array}}
\def\err{\end{array}}
\def\dsp{\displaystyle}
\newcommand{\rhobar}{\overline{\rho}}
\newcommand{\rhohat}{\hat{\rho}}
\newcommand{\xibar}{\overline{\xi}}
\newcommand{\deltabar}{\overline{\delta}}
\newcommand{\sigmabar}{\overline{\sigma}}
\newcommand{\deltahat}{\hat{\delta}}
\newcommand{\sigmahat}{\hat{\sigma}}
\def\Or{{\cal O}}
\def\La{{\cal L}}
\def\dotD {{\dot{D}}} 
\def\dotF {{\dot{F}}} 
\def\ddotF {{\ddot{F}}}
\def\ddotR {{\ddot{R}}}
\def\dotdel {{\ddot{\delta}}}
\def\ddotdel {{\ddot{\delta}}} 
\def\xdot {{\dot{\bf x }} }
\def\dota {{\dot{a}}} 
\def\vpsi {{\bf \Psi }}

\font\twelveBF=cmmib10 scaled 1200
\newcommand{\bte}{\hbox{\twelveBF $\theta$}}
\newcommand{\x}{\hbox{\twelveBF x}}
\newcommand{\q}{\hbox{\twelveBF q}}
\newcommand{\vv}{\hbox{\twelveBF v}}
\newcommand{\y}{\hbox{\twelveBF y}}
\newcommand{\r}{\hbox{\twelveBF r}}
\newcommand{\k}{\hbox{\twelveBF k}}
\newcommand{\lexp}{\mathop{\bigl\langle}}
\newcommand{\rexp}{\mathop{\bigr\rangle}}
\newcommand{\rexpc}{\mathop{\bigr\rangle_c}{}}
\newcommand{\eq}{{equation~}}

\begin{abstract}

In Part I of this series, we introduced the Spherical Collapse (SC) 
approximation in Lagrangian space as a way of estimating the cumulants 
$\xi_J$ of density fluctuations in  cosmological Perturbation Theory (PT).
Within this approximation, the dynamics is decoupled from the statistics 
of the initial conditions, so we are able to present here the cumulants
for generic Non-Gaussian initial conditions, which can be estimated to 
arbitrary order including the smoothing effects. The SC model turns out 
to recover the exact leading-order non-linear contributions 
up to terms involving 
non-local integrals of the  $J$-point functions. We argue that for the 
hierarchical ratios $S_J$, these non-local terms are sub-dominant and 
tend to compensate each other. 
The resulting predictions show a non-trivial time evolution that can be
used to discriminate between models of structure formation. We compare 
these analytic results to Non-Gaussian N-body simulations, which turn 
out to be in very good agreement up to scales where $\sigma \simlt 1$.  

\end{abstract}






\section{Introduction}

The large scale galaxy distribution can be used to study 
the origin and dynamics of cosmological fluctuations.
The one-point statistical clustering of matter density fluctuations 
$\delta_R$, smoothed over scale $R$, is characterized here
in terms of the reduced $J$th-order moments 
$\xibar_J(R) \equiv {\lexp \delta^J_R \rexpc}$.
Assuming that gravity is the dominant dynamical effect,
the evolution of $\xibar_J(R,t)$ is completely 
fixed by the initial conditions (IC): $\xibar_J(0,R)$.
The problem is that even in the simplest case of pressureless (collisionless)
matter, we do not have the exact solution to the dynamical equations.
Here we concentrate on the approximation given by perturbation theory (PT)
which  provides a framework to study small departures
from the linear theory solution.
The question we want to address is: what are the predictions
of PT for a given set of (Non-Gaussian) IC? 
This is important because large scale galaxy surveys
can then be used to discriminate between models of structure formation
(\eg Silk \& Juszkiewicz 1991).

We already have a good idea of how to answer this
question for Gaussian initial conditions (GIC), both at leading order, 
or tree-level (Peebles 1980, Fry 1984, Bernardeau 1992), and also
with higher-order (loop) corrections (Scoccimarro \& Frieman 1996a, 1996b,
Fosalba \& Gazta\~naga 1998).
Comparison with observations and simulations made it necessary
to develop PT also for the smoothed fields 
(\eg Goroff \etal 1986, Juszkiewicz \etal 1993, 
Bernardeau 1994a, 1994b).

The analytic calculations  in the literature for Non-Gaussian 
initial conditions (NGIC) refer to the 
variance, $\xi_2$,  skewness $S_3$,
 (see Fry \& Scherrer 1994, FS94 hereafter) 
and the kurtosis, $S_4$,
(see Chodorowsky \& Bouchet 1996, CB96 hereafter). These studies
are important from the theoretical point of view,
but of little help in practice because they
only apply to the {\em unsmoothed} fields. 
Numerical work by Weinberg \& Cole (1992)
provided interesting insights on different aspects of some particular 
Non-Gaussian models  but did not address the general question 
of the initial conditions (other works include Moscardini \etal 1993).
Gazta\~{n}aga and M\"{a}h\"{o}nen (1996), studied the case of
strongly NGIC in N-body simulations as compared to GIC, showing
important differences that could be used to discriminate models with
current observations.  

Here we follow the path set up in Paper I (Fosalba \& Gazta\~naga 1998)
and study the case of
a generic Non-Gaussian model for the IC, under the spherical collapse
(SC) approximation, which will allow us to present predictions
for the smoothed fields.
In \S2 we set the problem of NGIC and recall
the SC model. In \S3 we present the predictions for a generic
case of NGIC and compare them to N-body simulations and 
some previous calculations.
We end up with a discussion and the final conclusions in \S4.

\section{Non-Gaussian Initial Conditions}

For a given clustering in the initial conditions, we would like
to derive the final clustering using
the exact dynamics that rule the evolution of the underlying field.
To characterize the clustering,
we concentrate on the study the statistical properties
of one-point $J$th-order  moments $m_J \equiv <\delta^J>$ of smoothed density
fluctuations $\delta$. 
In particular we will use a combination of moments,
the so call {\it connected moments} or {\it reduced  cumulants} 
$\xibar_J$,  which  of a given $J$
carry statistical information independent of the
lower order moments, and are formally denoted by a bracket with
subscript $c$. These cumulants can be easily obtained  by using the
generating function method:
\beq
\xibar_J = \lexp\delta^J \rexpc = \left.{d^J \over{dt^J}} 
\ln \lexp e^{t \delta} \rexp\right|_{{t=0}}. 
\label{generating}
\eeq
A {\it dimensional scaling} of the higher-order moments in terms of the second
order one $\xibar_2=\sigma^2$ is given by the following ratios:
\beq
B_J \equiv {\xibar_J \over{\sigma^J}} = {\xibar_J \over{\xibar_2^{J/2}}}.
\label{dimratios}
\eeq
For Gaussian initial conditions (GIC) in PT, 
it is more useful to introduce the {\it hierarchical coefficients},
\beq
S_{J}\,=\,\frac{\xibar_J}{{\xibar_2^{J-1}}} = B_J \,{[\xibar_2]^{1-J/2}}
\label{sj}
\eeq
as PT predicts them to be time-independent quantities on 
large scales, for GIC.
These amplitudes are also called normalized one-point cumulants or 
reduced cumulants.
We shall also use
{\it skewness}, for $S_3$ and {\it kurtosis}, for $S_4$. 

As in Paper I, we start by 
 expanding the density 
contrast $\delta$ assuming that it is small, as is usually done in the 
context of perturbation theory (PT) in Euler space,
\beq
\delta(\x,t) = \delta_1(\x,t) 
+ \delta_2(\x,t)+ \delta_3(\x,t) ..., 
\label{pt}
\eeq
were we consider $\delta_n << \delta_{n-1}$.
The first term $\delta_1 \equiv \delta_l$
is the solution to the linearized field equations. The second term
$\delta_2 \sim \delta_l^2$ is the 
second-order solution, obtained by using the linear solution in the 
source terms, and so on.

In Fourier space, we write, 
$\delta (\k) \equiv \sum_{n=1}^{\infty} \delta_n (\k)$, being 
$\delta_n (\k)$ the $n$-th order perturbative contribution. The latter
is expressed as an $n$-dimensional integral over the kernels, $F_n$, 
that encode the 
nonlinear coupling of modes due to the gravitational evolution,
\bea
\delta_n(\k) &=& \int  d^3q_1 ... d^3q_1 \, \delta_D(\k-\q_1...-\q_n) \, \times
\nn \\
&\times& F_n(\q_1...\q_n) 
\, \delta_1(\q_1)...\delta_1(\q_n)
\label{deltan}
\eea 
where $\delta_D$ is the
Dirac function and the kernels 
$F_n$ are given by symmetric homogeneous functions of $\q_n$
with degree zero, that is, some geometrical average (see Fry 1984, 
Goroff \etal 1986, Jain \& Bertschinger 1994,
Scoccimarro \& Frieman 1996a). 

\subsection{Linear Theory}

If we only consider the first term in the PT series, Eq.[\ref{pt}],
and the growing mode, 
the cumulants of the evolved field will just be:
\beq
\xibar_J \equiv \lexp \delta^J \rexpc =  \lexp \delta_l^J \rexpc =
D^J ~ \lexp \delta_0^J \rexpc,
\eeq
where $D$ is the linear growth factor and 
$\lexp \delta_0^J \rexpc$ correspond to the cumulants of the IC.
Consistently, the hierarchical ratios (see Eq.[\ref{sj}]) will scale as:
\beq
S_J = {S_J(0)/{D^{J-2}}},
\eeq
were $S_J(0)$ are the initial ratios.
Note that this implies that the linear growth erases the initial
hierarchical ratios, so that $S_J \rightarrow 0$, as time evolves (and $D
\rightarrow \infty$). 

In terms of the dimensional scaling, see Eq.[\ref{dimratios}], we have,
$B_J = B_J(0)$, so that the linear growth preserves the initial values.
Note that if we want to do a meaningful calculation of these
ratios or the cumulants, in general, we might need to consider more
terms in the perturbation series, Eq.[\ref{pt}], depending on 
the statistical properties of the IC, \eg how they scale
with the initial variance, which is typically the smallness
parameter in the expansion of the cumulants.

For GIC both $B_J(0) = S_J(0) = 0$,
and we have to consider higher-order terms in the perturbation
series to be able to make a non-vanishing prediction.

\subsection{Gaussian Tree-Level}
\label{sec:gausstree}

The computation of the cumulants in PT dates back to Peebles (1980) work
where the leading order contribution to the skewness was obtained making
use of the second-order PT, \eg $F_2$.
Fry 1984 extended Peebles analysis by making the connection between tree
diagrams (or tree-graphs)
and the perturbative contributions to leading order in the 
GIC case. With the help of this formalism
he was able to obtain the leading order contributions for the 
three- and four-point 
functions making use of the 2nd and 3rd order kernels, $F_2$ and $F_3$, in PT. 
Furthermore, Fry found 
that, in general,  the lowest order (tree-level)
connected part that contributes to $\xi_J$ are of order $2(J-1)$ in $\delta_1$.
Note that this involves the cancellation of $J-2$ contribution to the 
moment of order $J$, $m_J$. This is a property of the GIC
for which all $\lexp\delta_1^J \rexpc$ vanish 
for $J>2$.  For GIC the leading order contribution is only given by 
tree-graphs and is therefore called the 'tree-level' (see also Paper I).

Bernardeau (1992) found the generating function of the one-point cumulants 
to leading order for GIC. Here (and in Paper I) we present a simpler
derivation, inspired in Bernardeau's work, that can be extended to 
higher-order (loop) corrections and NGIC.

\subsection{The Monopole Approximation \&
The Spherical Collapse Model}

Given the kernels $F_n$ in Eq.[\ref{deltan}],
we define the {\em monopole} contribution to $F_n$
as the spherically symmetric (angle) average:
\beq
F_n^{l=0} \equiv <F_n> = c_n/n!.
\label{monop}
\eeq
The monopole approximation, $\delta$ is the one resulting by substituting
$F_n$ by $<F_n>$. Under this approximation, we obviously have,
from Eq.[\ref{deltan}]:
\beq
\delta_n (\k) = {c_n \over n!}\,\delta_l (\k) * \dots * \delta_l (\k),
\eeq
where $*$ means convolution (in Fourier space),
so that, in real space
$\delta (\k) \equiv \sum_{n=1}^{\infty} \delta_n (\k)$ becomes:
\beq
\delta (\x) \equiv \sum_{n=1}^{\infty}\,\delta_n (\x) = 
f[\delta_1(\x)]=
\sum_{n=1}^{\infty}\,{c_n \over n!}\,[\delta_l(\x)]^n.
\label{local} 
\eeq 
Thus, the
monopole contribution 
to the cumulants in PT is given by a local-density
transformation Eq.[\ref{local}],
whose coefficients $c_n$ 
are to be determined by the kernels $F_n$, which are found
by solving the perturbative equations under the 
relevant dynamics (in Fourier space). 

We can now easily estimate 
{\it all} the 1-point statistical properties in the {\em monopole} 
approximation to PT.
This can be done by using the generating function method Eq.[\ref{generating}],
with the field $\delta$ given by Eq.[\ref{local}].
When the initial conditions are hierarchical, the resulting expressions
can be found in Fry \& Gazta\~naga (1993), who consider a generic 
local-density transformation and find, to leading terms in $\sigma_l$:
\bea
S_{3} &=& S_3^{IC} + 3 c_2 +\Or(\sigma_l^2) \nn \\
S_{4} &=& S_4^{IC} + 12 c_2 S_3^{IC} + 4 c_3 + 12 c_2^2 +\Or(\sigma_l^2) \nn \\
S_{5} &=& S_5^{IC} + 20 c_2 S_4^{IC} + 15 c_2 {S_3^{IC}}^2 + 
        (30c_3 + 120 c_2^2) S_3^{IC}  \nn \\
&+& 5 c_4 + 60 c_3 c_2+ 60 c_2^3 +\Or(\sigma_l^2)
\label{localh}
\eea
where $S_J^{IC}$ are the hierarchical amplitudes in linear theory
(which are given by the IC).
If the IC are not hierarchical one might have to consider
more terms to estimate the leading order (see \S\ref{sec:ngic}). 
These arguments are valid for any dynamics and
also for both Euler or Lagrangian space, they apply to any 
leading order calculation where Eq.[\ref{deltan}] is valid.

To estimate the PT contribution to the cumulants (\eg $c_n$) it is  not
necessary to calculate the structure of the kernels $F_n$ from the exact solution
to the field equations of $\delta$, as in Goroff \etal (1986).
Given the equations for the evolution of the field 
one can determine Eq.[\ref{local}] and therefore $c_n$
by just requiring the solutions to be {\em spherically symmetric}.
In Paper I, it was shown that for gravity, the spherically symmetric
solution to the dynamical equations 
is given by the Spherical Collapse (SC), whose
solution is well-known [\eg see \S4.1 in Paper I].
Thus, $c_n=\nu_n$, without need of estimating the kernels
$F_n$ or any integral. 
In particular, the $\nu_n$ coefficients are derived by taylor-expanding the
parametric solution to the SC model around $\delta_l = 0$
(see Paper I, \S 4.1).
The connection between the Gaussian tree-level in PT and the SC model was
already pointed out by Bernardeau (1992, 1994a, 1994b) although, there, 
the tree-level
amplitudes were derived using the rather complicated formalism of the 
{\em vertex generating function} instead of the density contrast itself.
 
For GIC we also showed in Paper I how the 
{\it monopole} approximation gives the exact leading order, which is
given by the tree-level solution. 
We next will argue that it also gives the exact contribution to the
tree-graphs that appear for NGIC.

\subsection{Cumulants for a Non-Gaussian Tree-Level}
\label{ngtree}

On estimating the PT predictions for cumulants $\xibar_J$,
Fry (1984) described the connection between tree
diagrams (or tree-graphs)
and the perturbative contributions to leading order in the 
Gaussian case.
These  Gaussian {\it tree-level} terms will also contribute
to the estimation $\xibar_J$ 
in the case of NGIC, as they come from the products of 
linear terms, $\delta_1$,
induced by the two-point function.
But  there are additional terms in the Non-Gaussian case.
For example, consider the kurtosis:
\beq
\lexp \delta^4 \rexp = \lexp ( \delta_1 + \delta_2 + \dots)^4 \rexp
= \lexp \delta_1^4 \rexp + 4 \lexp \delta_1^3\delta_2\rexp + \dots.
\label{kurt}
\eeq
Now,  $\delta_2$ is given by the 2nd order kernel $F_2$ in Eq.[\ref{deltan}]:
$\delta_2 \sim F_2 \delta_1^2$  so that
the term $\lexp \delta_1^3\delta_2\rexp$ above involves an integral
over $F_2\lexp \delta_1^5 \rexp$, which is zero in the Gaussian case, but
gives in general:
\beq
\lexp \delta_1^5 \rexp =
\lexp \delta_1^5 \rexpc + 10 \lexp \delta_1^3 \rexpc \lexp \delta_1^2 \rexp.  
\label{treeng}
\eeq
Consider first the second term of the above equation. The only connected part 
of the above term that contributes to the $F_2\lexp \delta_1^5 \rexp$  integral
in the kurtosis, is just given by a tree-graph, connected with no loops, of the type:
\bea
&& \int  d^3q_1 d^3q_2 \, \delta_D(\k_1-\q_1-\q_2) \, F_2(\q_1,\q_2) \nn \\ 
&& \lexp \delta_1(\q_1)\delta_1(\k_2) \delta_1(\k_3) \rexp \lexp
\delta_1(\q_2)\delta_1(\k_4) \rexp,
\label{deltani2}
\eea 
here again,
like in the Gaussian case [Eq.[23] in Paper I],
the last two factors do not have any dependence on the angle,
$\q_1\q_2$,  and the 
contribution of the kernel $F_2$ is  just given by the
geometric average, \eg  the number $c_2/2$.
This illustrates that 
{\em the local transformation, Eq.[\ref{local}], given by the monopole, 
also accounts for the exact contribution from
the tree-graphs in the Non-Gaussian case},
\eg the contribution $ 12 c_2 S_3^{IC}$ 
to $S_4$ in  the local expression (\ref{localh}). It also 
accounts for the direct terms from linear theory [first terms to the
right hand of equations (\ref{localh})], which are zero in the Gaussian
case, and are not regarded as tree-graphs.


Nevertheless, for Non-Gaussian IC the tree-graphs
do not necessarily include all the leading-order contributions,
unlike the Gaussian case. There are terms
like $\lexp \delta_1^5 \rexpc$ in Eq.[\ref{treeng}], which
also comes from $\lexp \delta_1^3\delta_2\rexp$ and contribute
to the kurtosis
Eq.[\ref{kurt}].
The contribution involves an integral of $F_2 \lexp \delta_1^5 \rexpc$.
In the monopole
approximation, Eq.[\ref{local}], this term is simply given by 
$c_2/2  \lexp \delta_1^5 \rexpc$. While
in the exact calculation, its value could have contributions
from higher multipoles, depending on the particular form of
the initial conditions $\lexp \delta_1^5 \rexpc$, which 
is needed to do the integration. FS94  have estimated this multipole 
integration
for several Non-Gaussian models for the initial kurtosis, $\xi_4^{l}
= \lexp \delta_1^4 \rexpc$.
They typically find (see their Table 1) that the coefficients
$I[\xi_4^{l}]/\xi_4^{l}$, which corresponds to the relative
contribution of the higher multipoles,
are smaller than unity, thus typically smaller than the monopole
contribution $c_n/n! \simgt 1$ (see below Eq.[\ref{eq:exacts3}]).
This is not surprising as correlations typically decrease with distance,
thus reducing the contribution from non-spherical geometries.
Thus, in general, on expects the monopole to be the dominant contribution.
In the case of hierarchical initial conditions,
$\lexp \delta_1^5 \rexpc$ is of order $\sigma_l^8$,
while $\lexp \delta_1^3 \rexpc \lexp \delta_1^2 \rexp$ is
of order $\sigma_l^6$. Thus in this case the monopole (local)
approximation {\em exactly} accounts  for all the leading-order terms.

\subsection{Smoothing Effects}

In \S 4.4 of Paper I, it was shown how to 
relate the {\it smoothed}
cumulants of the evolved distribution with the {\it smoothed}
cumulants in the initial one. The arguments presented there were
general and could be applied to both Gaussian and Non-Gaussian
initial conditions. In the case of a 
power-law power spectrum $P(k) \sim k^{n}$, the smoothed variance
is also a power-law $\sigmahat_l \sim R^{\gamma/2}$, where
$\gamma= -(n+3)$. We then have:
\beq
\deltahat[\deltahat_l] \sim f[\deltahat_l  (1+\deltahat)^{\gamma/6}]
\label{deltahat2}
\eeq
up to a normalization factor given by Eq.[39] in Paper I.
Note that this final result as well as the general expression
agrees with Bernardeau (1994a) arguments,
based on the vertex generating function, 
but they do not limit us to Gaussian 
initial conditions or the leading-order term. 
Here again, the
vertex generating function ${\cal G}(-\tau)$, corresponds to 
cumulants in Euler space, while our local-density relation
$f(\delta_l)$, applies to Lagrangian space. To leading order,
they both  give identical results for Gaussian IC, but 
they yield different results in general for Non-Gaussian IC or
for higher-order terms with Gaussian IC.

\section{Predictions for Non-Gaussian Initial Conditions}
\label{sec:ngic}

Here we study the more general case of 
evolution in PT
from Non-Gaussian initial conditions (NGIC), where $\xibar_J(0) \neq 0$.
If we use $\delta_n \sim \delta_1^n$, the first 
perturbative contributions are:
\bea
\xibar_J(t) 
& \simeq &  D^J~\xibar_J(0) \,+\, J/2~d_2~D^{J+1}~\xibar_{J+1}(0) + \dots \nn \\
&+&  D^{2(J-1)}~S_J^G~ \xibar_2^{J-1}(0)  + \dots
\label{nongaus}
\eea
where $S_J^G \equiv S_{J,0}^G$ (see notation in \S\ref{sec:scm}) 
are the Gaussian tree-level amplitudes,
\bea
S^G_3 &=&  {34\over 7} + \gamma \nn \\
S^G_4  &=& {60712\over 1323} + {62\over 3}\,\gamma + 
{7\over 3}\,\gamma^2   
\eea
and so on, for a power-law power spectrum 
and a top-hat window (see also \eg \S 5 in Paper I.)
These amplitudes are intrinsically
gravitational as the GIC ones are zero for $J>2$.

In the case of NGIC, it is more difficult to present a perturbation series
because it depends on how $\xibar_J(0)$ scales with $\xibar_2(0)$.
In a statistically homogeneous distribution,
 $\xibar_J(0)  \rightarrow 0$ in the limit
 $\xibar_2(0) \rightarrow 0$, so that we can write
 $\xibar_J(0)  \rightarrow [\xibar_2(0)]^{\alpha}$,
with $\alpha=\alpha[J]$. So we may consider the general case:
\beq
\xibar_J(0) = B_J \,\left[\xibar_2(0)\right]^{\alpha[J]}.
\eeq
Let us consider the different possibilities for $\alpha=\alpha[J]$
in the above series (\ref{nongaus}).
When $ \alpha > J-1  $ the initial conditions are {\it forgotten},
as the leading-order effect of the evolution is still the 
hierarchical term, $S_J^G$, which dominates the evolution.
In the more general case, the contribution to $\xibar_J$ from 
Non-Gaussian initial correlations of order larger than $J$ are
suppressed by powers of the growth factor $D$ and they only become
important at late times.
When $ J/2 < \alpha < J-1 $ we have quasi-Gaussian 
but non-hierarchical initial conditions.
The evolution in $ \xibar_J$  has
a dominant non-hierarchical term that grows as $D^J$, while the 
hierarchical term grows as 
$D^{2(J-1)}$ and may not become significant until $ \xibar_2 \sim 1 $.
Note that, as pointed out by FS94, there is an
additional Non-Gaussian term that grows as $D^{J+1}$ and,
for the skewness ($J=3$), it contributes directly to $S_3 \equiv
\xibar_3/\xibar_2^2$ so that $S_3 \neq S_3^{G}$ at all times.
When $ \alpha = J/2$, as in the dimensional scaling where 
$\xibar_J(0) \simeq \xibar_2(0)^{J/2}$,
both terms that grow as $D^4$ depend on
$\xibar_2^2$ and their total amplitude is  
$S_3^{G}+3/2 d_2$.
If $ \alpha < J/2 $  there are strongly NGIC
that dominate the evolution
as far as $\xibar_2$ is small (\ie on large scales). 

Here we will concentrate on the  $ \alpha = J/2$ case, the
transition to the  strongly NGIC.
For the dimensional scaling which typically arises in topological defects
models such as textures, we have,
\beq
\xibar_J \, = \, B_J \,\xibar_2^{J/2},
\eeq
where the $B_J$ amplitudes are constants and have been predicted to 
be of order unity
(see Turok and Spergel 1991, Gazta\~{n}aga and M\"{a}h\"{o}nen 1996).

\subsection{The Spherical Collapse Model Results}
\label{sec:scm}

In the SC model, 
it is straightforward to work out the perturbation 
expansion. We can proceed as in Eq.[\ref{localh}] by just 
keeping track of the dominant terms, as we do not need to solve
any additional equations. We represent the different orders in the expansion
of the cumulants
following the notation introduced in Paper I: 
\bea
\sigma^2 &=& \sum_{i} {s_{2,i}\,\sigma_l^i} = \sigma^2_l + s_{2,3}\,
\sigma^3_l \,+\, s_{2,4}\,\sigma^4_l + \cdots
\\
S_J &=& \sum_{i} {S_{J,i}\,\sigma_l^i} = \cdots  \,+
S_{J,-1}\,\sigma_l^{-1}+\,S_{J,0} \nn \\
&+& S_{J,1}\,\sigma_l^1+\cdots,
\eea
and we find for the first non-vanishing perturbative 
contributions:
\bea
s_{2,3} &=&  
\left[{S_3^G\over 3}-1\right]\,B_3 \nn \\
s_{2,4} &=&  
3 - {4\over 3}\,S_3^G - {5 \over 18}\,\left(S_3^G\right)^2 + 
{S_4^G\over 4}   \nn \\
&+& \left[1 - {S_3^G\over 2} - 
        {\left(S_3^G\right)^2\over 12} + {S_4^G\over 12}\right]\,B_4 
\nn \\	
S_{3,-1} &\equiv& S_3^{(0)} = {B_3} \nn \\
S_{3,0} &=&
 S_3^G - 2\,\left[{S_3^G\over 3}-1\right]\,B_3^2 
+  \left[{S_3^G\over 2}\,-1\right]\,B_4\, \nn \\
S_{3,1} &=& 
\left[{S_3^G\over 6} -{17 \over 18}\,\left(S_3^G\right)^2 
+ {5\over 8}\,S_4^G \right]\,B_3 \nn \\
&+ & \left[3\,-  2\,S_3^G + {\left(S_3^G\right)^2\over 3} \right]\,B_3^3 
\nn \\
&+& \left[- 4 + 
{8\over 3}\,S_3^G - 
 {\left(S_3^G\right)^2\over 6} - {S_4^G\over 6} \right]\,B_3\,B_4 
 \nn \\
&+&   
\left[1 - {2\over 3}\,S_3^G - 
        {\left(S_3^G\right)^2 \over 12} + {S_4^G\over 8} \right]\,B_5 \nn \\
S_{4,-2} &\equiv& S_4^{(0)} = {B_4} \nn \\
S_{4,-1} &=& 
4\,S_3^G\,B_3 + 
\left[3 - S_3^G\right]\,B_3\,B_4 + \left[{2\over 3}\,S_3^G - 1\right]\,B_5 
\nn \\
S_{4,0} &=& S_4^G +  \left[3 + 7\,S_3^G - 
{14\over 3}\,\left(S_3^G\right)^2 + 
     {3\over 2}\,S_4^G \right]\,B_3^2    \nn \\
&+&  \left[-1 - {10\over 3}\,S_3^G + 
{\left(S_3^G\right)^2 \over 6} +    
     {5\over 4}\,S_4^G \right]\,B_4 \nn \\
&+& \left[6 - 4\,S_3^G 
   +  {2\over 3}\,\left(S_3^G\right)^2  \right]\,B_3^2\,B_4 \nn \\ 
&+& \left[- 3\, + {3\over 2}\,S_3^G + 
{\left(S_3^G\right)^2 \over 4} - {S_4^G\over 4} \right]\,B_4^2  
\nn \\
&+& \left[- 3\, +  
3\,S_3^G\, - 
     {2\over 3}\,\left(S_3^G\right)^2  \right]\,B_3\,B_5 \nn \\ 
&+& \left[1\, - {5\over 6}\,S_3^G - 
{\left(S_3^G\right)^2 \over 18} + {S_4^G \over 6} \right]\,B_6 
\nn \\
S_{5,-3} &\equiv & S_5^{(0)} = {B_5} \nn \\
S_{5,-2} &=&  
5\,S_3^G\,B_3^2 + {20\over 3}\,
S_3^G\,B_4 - 
        4\,\left[{S_3^G \over 3}-1\right]\,B_3\,B_5   \nn \\
&+& \left[  
{5\over 6}\,S_3^G \,-1\right]\,B_6 
\nn \\
S_{6,-4} & \equiv & S_6^{(0)} = {B_6} \nn \\ 
S_{6,-3} &=&
20\,S_3^G\,B_3 \,B_4  
+ 10\,S_3^G\,B_5 
- 5\,\left[{S_3^G \over 3}-1 \right] \,B_3 \,B_6  \nn \\
&+& \left[S_3^G\,-1\right]\,B_7.
\label{sj_dim}
\eea
We have chosen to write the expressions as a function of
$S_J^G$, the tree-level hierarchical coefficients
for GIC, so that when $B_J=0$, one immediately recovers 
the Gaussian results.
This is possible because, as mentioned in previous
sections, the Gaussian tree-level contains all the information of
the SC model (see \eg Appendix A1 in Paper I).
Note that this includes both the smoothed and the unsmoothed
case, by just replacing the correct value of $n$ or $\gamma_p$
in $S_J^G$.
We stress that, unlike the case of GIC, odd powers in the linear variance
also contribute to the cumulants for generic NGIC as already pointed out
in recent papers (see FS94, CB96).

\begin{table}

\begin{center}

\begin{tabular}{|c||c|c|c|c|}
\hline \hline
SC & Unsmoothed & \multicolumn{3}{c|}{Smoothed} \\ 
\hline \hline
NGIC& $\gamma=0$ & $\gamma=-1$ & $\gamma=-2$ & $\gamma=-3$ \\ \hline
$B_J=1$& $n=-3$ & $n=-2$ & $n=-1$ & $n=0$  \\ 
\hline \hline
$s_{2,3}$   & 0.62 & 0.29 & -0.05 &  -0.38 \\ 
\hline
$s_{2,4}$   & 1.87 & 0.74 & 0.44 &  0.98 \\
\hline
$s_{2,5}$   & 3.36 & 0.60 & -0.05 &  -1.05 \\ 
\hline \hline
$S_{3,0}$ & 5.05 & 4.21 & 3.38 & 2.55 \\ 
\hline
$S_{3,1}$ & 7.26 & 3.91 & 1.55 & 0.19 \\ 
\hline
$S_{3,2}$ & 23.53 & 7.37 & 1.18 & 0.20  \\ 
\hline \hline
$S_{4,-1}$ & 19.81 & 16.14 & 12.48 & 8.81  \\ 
\hline
$S_{4,0}$ & 85.88 & 52.84 & 28.31 & 12.27 \\ 
\hline
$S_{4,1}$ & 332.51 & 128.51 & 32.83 & 2.70 \\ 
\hline \hline

\end{tabular}

\caption[junk]{Values of the higher-order perturbative contributions in the
SC model from NGIC with $B_J=1$
 for the unsmoothed ($n=-3$) and 
smoothed ($n=-2,-1,0$) density fields for a top-hat window and a power-law 
power spectrum.}
\label{ngicsc}
\end{center}

\end{table}

Note that in Eq.[\ref{sj_dim}], there are two types of
contributions. First, there are terms that arise as a result of 
the non-linear gravitational
evolution alone ($\sim S_J^G$), which are not coupled to the IC, $B_J$ and
thus which contribute in the same way as for GIC.
Second, there are those terms coupled to the IC ($\sim B_J$) which
depend on the specific Non-Gaussian model for the IC. 

It is natural to introduce the non-linear {\em dimensional} ratios
for the texture-like NGIC (see Eq.[\ref{dimratios}]), 
\bea
B^{NL}_J &=& \sum_{i} B_{J,i}\,\sigma_l^i =
B_{J,0} + B_{J,1}\,\sigma_l^1 +\cdots,  
\eea
Notice that the linear term ($\sim {\cal O (\sigma_l^0)}$) remains  
equal to its initial value (contrary to the leading order to the
$S_J$ ratios, see Eq.[\ref{sj}]), while the $\sigma$-corrections 
(\ie the ${\cal O (\sigma_l)}$ terms) only appear when  
non-linear gravitational evolution sets up. In particular,
we have to keep $J-2$ corrective terms in order to include the
purely gravitational term (not coupled to the IC), $\sim S_J^G$.
Note that one has to include this hierarchical term to get an 
accurate estimate of the $B^{NL}_J$, 
as its contribution increases with decreasing scale
(see Figs \ref{s3textpt} \& \ref{s4textpt} for $S_J = B_J \sigma^{2-J}$).   
\bea
B_{3,1} &=& S_3^G + 
{1\over 2} \left(3 - S_3^G \right) B_3^2 + 
\left({S_3^G \over 2} -1 \right) B_4   \nn \\
B_{4,1} &=& 4 S_3^G B_3 + 2 \left(1 - {S_3^G \over 3}
\right) B_3 B_4 + \left({2\over 3} S_3^G -1 \right) B_5 \nn \\
B_{4,2} &=& S_4^G + 3 \left[1 + S_3^G - {10\over 9} \left(S_3^G
\right)^2 +
{S_4^G \over 2} \right] B_3^2 \nn \\
&+& \left[2 - {14\over 3} S_3^G - {1\over 9} \left(S_3^G
\right)^2 + {3 \over 2} S_4^G 
\right] B_4 \nn \\
&+& \left[3 - 2 S_3^G - {1\over 3} \left(S_3^G \right)^2 
\right] B_3^2 B_4 \nn \\
&+& \left[-2 + S_3^G + {1\over 6} \left(S_3^G \right)^2 
- {1 \over 6} S_4 \right] B_4^2 \nn \\
&-& 2 \left[-1 + S_3^G - {2\over 9} \left(S_3^G \right)^2 \right] B_3 B_5  
\nn \\
&+& \left[1 - {5\over 6} S_3^G - {1\over 18} \left(S_3^G \right)^2 + 
{1\over 6} S_4 \right] B_6   
\label{dim_rat}
\eea
Notice that for most of the cosmological models, 
quasi-linear scales (where PT applies)
have an {\em effective} spectral index $n_{eff}$, associated to a power-law 
power spectrum, $P(k) \sim k^{n_{eff}}$, 
within the range $n_{eff} \in [-1,-2]$, for which 
$S_3^G \approx 3$. This means that, for most of the models, the  
second term in $B_{3,1}$ \& $B_{4,1}$ have a negligible 
contribution.
     
Table \ref{ngicsc} shows the results for the NGIC 
for $B_J =1$ and different power indexes. These are to be
compared with the GIC case in Table 2 of Paper I.

\begin{figure}[t]
\centering
\centerline{\epsfysize=8.truecm 
\epsfbox{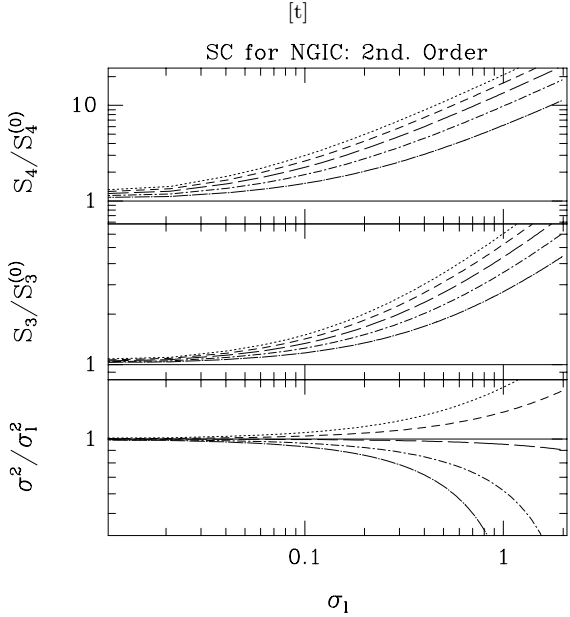}}
\caption[junk]{Departures from the tree-level 
contributions as the linear {\em rms} fluctuation grows, for the variance, 
skewness and kurtosis of the density field
as predicted by the SC model up to the 2nd non-vanishing 
perturbative contribution (first corrective term) for 
different values of the spectral index: the dotted line shows the $n=-3$
(unsmoothed) case, and the short-dashed ($n=-2$), long-dashed ($n=-1$),
dot short-dashed ($n=0$), dot long-dashed ($n=1$) depict the 
behavior for the smoothed density field. 
The solid line shows the tree-level values (or linear term
for the variance) as a reference.
It is shown the case of Non-Gaussian initial conditions with $B_J \approx 1$.}
\label{scvpttop1}
\end{figure}

\begin{figure}[t]
\centering
\centerline{\epsfysize=8.truecm 
\epsfbox{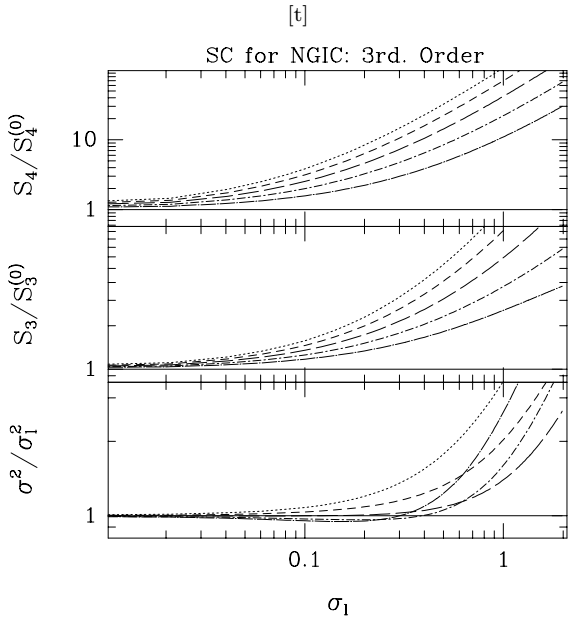}}
\caption[junk]{Same as Fig \ref{scvpttop1}, when the 
3rd perturbative contribution is taken into account.}
\label{scvpttop2}
\end{figure}

Figs \ref{scvpttop1} and \ref{scvpttop2} show the departures from the
tree-level amplitudes for the variance, skewness and kurtosis up to
the 2nd and 3rd perturbative order respectively, as the linear
rms fluctuation grows. It is seen how the 2nd order contribution
(first corrective term) for the variance exhibits a cancellation of
non-linearities for $n \approx -1$ which tends to disappear when the linear
rms density fluctuation approaches unity, once the next 
perturbative
order (second corrective term) is included, as shown in Fig \ref{scvpttop2}. 
We interpret that this is due to the fact that,
here, unlike the case for GIC,
the initial conditions still dominate
the 2nd perturbative contribution in the evolved variance and 
further (3rd order)
contributions change this picture since gravitational evolution takes over the
initial conditions. Put it another way, it takes longer for the 
gravitational evolution
to erase the trace of strongly NGIC.

On the other hand, for the skewness
and kurtosis we see a significant change from the Gaussian case that holds 
in the 2nd and 3rd order analysis: strong non-linear effects are
found for any value of the spectral index, contrary to what happened
for GIC where there was a characteristic index $n \simeq 0$, where
non-linear effects vanished. This difference must be due to the strong
Non-Gaussian character of the initial conditions. In fact, by looking at the
expressions for $S_3$ and $S_4$ for dimensional NGIC 
(see \eg Eq.[\ref{sj_dim}] above) we
see that, unlike the case of the variance, the higher-perturbative orders are 
completely coupled to the initial conditions ($\sim B_J$ terms), so that,
even for the smallest scales, deviations from the PT predictions for Gaussian
initial conditions (where gravitational evolution clearly dominates) must
show up. This deviation from the Gaussian prediction on small scales is
realized as a shift from the Gaussian values in the 
hierarchical contributions (order zero in $\sigma$, $S_{J,0}$).

\subsection{Comparison with the Gaussian Case}

We have explored the non-linear corrections to the 1-point cumulants
of the density field, within the SC model, for Gaussian (in Paper I) and 
Non-Gaussian dimensional initial conditions separately.
The aim of this section is to provide a direct way of comparing
the predictions in both cases
so as to get a further insight on how
the choice of initial conditions changes the non-linear evolution.

To simplify the comparison, we shall assume that the 
$B_J$ coefficients (see \eg Eq.[\ref{dimratios}]), are all equal, which is 
roughly what is expected from analytic models for Topological defects. 
We denote $\beta \equiv B_J$ as the {\em Non-Gaussian strength}. 
While the results in Table 1 assume
$\beta=1$, we will now display the  non-linear corrections 
as a function of $\beta$. We shall concentrate on the coefficients
of even corrections in powers of $\sigma$
($s_{2,4}$, $S_{J,0}$), 
as the odd corrections vanish for GIC. Note 
nevertheless that these latter could be dominant contributions when studying
the overall non-linear effect for NGIC. Using Eq.[\ref{sj_dim}], we find
for top-hat smoothing and a power-law spectrum:
\bea
s_{2,4} &=& {1909\over 1323}\,+\,
{143\over 126}\,\gamma + 
{11\over 36}\,\gamma^2 + \left({1705\over 3969}\,+\,
{26\over 63}\,\gamma + 
{\gamma^2 \over 9} \right)\,\beta \nn \\
S_{3,0} &\equiv& S_3^{(0)} =  {34\over 7} + \gamma 
+ \left({10\over 7}\,+\,{\gamma \over 2} \right)\,\beta -
\left({26\over 21}\,+\,{2\over 3}\,\gamma \right)\,\beta^2 \nn \\
S_{4,0} &\equiv& S_4^{(0)} = {60712\over 1323} + {62\over 3}\,\gamma + 
{7\over 3}\,\gamma^2 \nn \\
&+& \left({188105\over 3969}\,+\,
{550\over 21}\,\gamma +{41\over 12}\,\gamma^2 \right)\,\beta  \nn \\
&-& \left({12841\over 1323}\,+\,
{253\over 21}\,\gamma +{13\over 6}\,\gamma^2 \right)\,\beta^2  \nn \\
&+& \left({338\over 147}\,+\,
{52\over 21}\,\gamma +{2\over 3}\,\gamma^2 \right)\,\beta^3.   
\eea
The case of GIC is reproduced by just setting $\beta=0$, while one
could expect $\beta \simeq 1$ for a defect model. Recall that
$\gamma= -(n+3)$.

\begin{figure}[t]
\centering
\centerline{\epsfysize=8.truecm 
\epsfbox{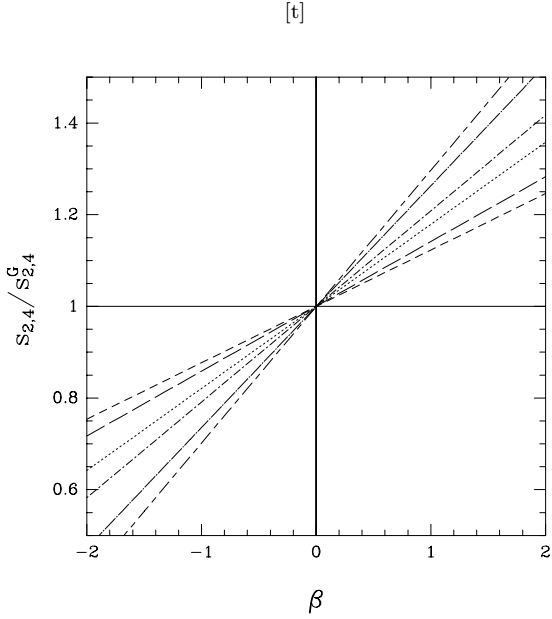}}
\caption[junk]{Departures from the Gaussian behavior in the coefficient 
$s_{2,4}$ of the $\sigma^4$ correction  to the linear variance
as a function of the Non-Gaussian strength $\beta = B_J$. 
The values displayed correspond 
to $n=-0.5$ (dotted line), $n=-1$ (short-dashed), $n=-1.5$ (long-dashed),
$n=-2$ (dot short-dashed), $n=-2.5$ (dot long-dashed), $n=-3$ 
(short-dashed long-dashed).}    
\label{s24ng}
\end{figure}

\begin{figure}
\centering
\centerline{\epsfysize=8.truecm 
\epsfbox{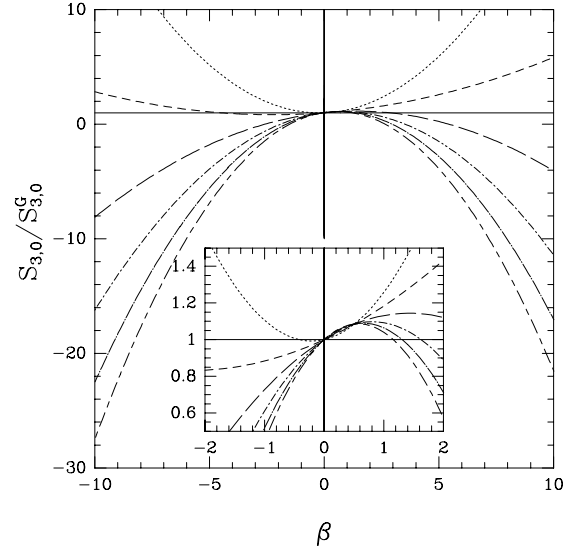}}
\caption[junk]{Same as Fig. \ref{s24ng} for the hierarchical 
contribution to the skewness, $S_{3,0}$. 
 The inset in the bottom shows a detail of the
plot for a shorter range of the Non-Gaussian parameter $\beta$.}
\label{s30ng}
\end{figure}

\begin{figure}
\centering
\centerline{\epsfysize=8.truecm 
\epsfbox{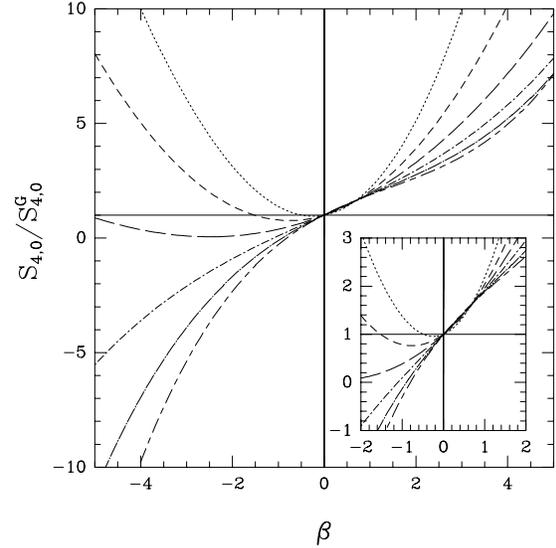}}
\caption[junk]{Same as Fig. \ref{s24ng}
for the hierarchical 
contribution to the kurtosis, $S_{4,0}$.} 
\label{s40ng}
\end{figure}

Figures \ref{s24ng}-\ref{s40ng} show the deviations from the
Gaussian perturbative contributions in the Non-Gaussian
model as a function of the {\em Non-Gaussian strength} $\beta$, for several
smoothing values, $n$.
They show the robustness of the arguments drawn 
from the $\beta = 1$ case (see Table 1).
The Non-Gaussian contributions are larger 
than the Gaussian one for $1 \simgt \beta \simgt 0$. 
For positive NGIC ($\beta >0$), 
this trend is typically enhanced when 
smoothing effects increase.

However, for negative values of the Non-Gaussian strength $\beta$, the
behavior changes. It is found that 
the variance falls below the Gaussian value, 
while the Non-Gaussian $S_J$ coefficients may 
overestimate or underestimate the Gaussian
prediction depending on the scales (effective spectral index) we
are looking at. Typically it underestimates the Gaussian value on 
quasi-linear scales.

The general behavior found for the variance ($s_{2,4}$)
seems to be in qualitative agreement with the conclusions drawn 
by FS94  which pointed out 
that a larger (lower) variance than the Gaussian one is expected for models
with positive (negative) initial skewness (see also Moscardini \etal 1993).
But note that this trend changes when $S_3^G<3$, since, then $s_{2,3}<0$, 
which is
the leading term in the variance (see Eq.[\ref{sj_dim}]). 
This was not detected
by FS94 because they did not include smoothing effects.
Thus, for $n > -8/7 \approx -1.14$ or $\gamma < -1.85$, 
there is a change in this
trend, with lower variance for more positive skewness. In observations
and also in CDM models (see Gazta\~naga \& Baugh 1998), 
the spectral index is $n \simlt -1$ in
the weakly non-linear scales, going to $n> -1$ on large scales.
This will produce a characteristic change in the
shape (see Figure \ref{x2text} below).

For $S_{3,0}$, 
a critical index is observed at $n =-8/7 \approx -1.14$ ($\gamma=-13/7$) 
below which
the Non-Gaussian hierarchical amplitude has a maximum as a function of
$\beta$ which depends on the value of $n$ (see inset in Fig \ref{s30ng}).
For $n > -8/7 \approx -1.14$ there is instead a minimum
as a function of $\beta$. 
These bounds  could be of interest when interpreting the 
estimations of $S_3$ from observations.
Given a value of the index $\gamma$,
there is a maximum ( $\gamma >-13/7$) or minimum ( $\gamma <-13/7$) in the 
non-linear coefficient at:
\beq
\beta_{m} \, = \, {{3\,(20+7\gamma)}\over{8\,(13+7\gamma)}}
\eeq
for which  the ratio is:
\beq
\left. {S_{3,0}\over{S_{3,0}^{G}}}\right\vert_{\beta_m} \, = \,
{{7\, (2192+1624\gamma+245\gamma^2) }\over{32\,(442+329\gamma+49\gamma^2)}}
\eeq


For $S_{4,0}$
we can also see some bounds (which depend on the spectral index) 
that could be of practical interest.

\subsection{Comparison with N-body Results}
\label{ng_nbody}

\begin{figure} 
\centering
\centerline{\epsfysize=8.truecm 
\epsfbox{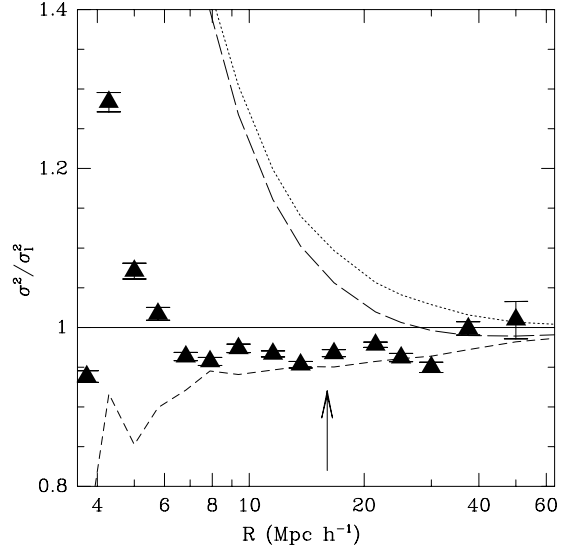}}
\caption[junk]{The ratio of the non-linear to the linear 
variance $\sigma^2/\sigma_l^2$
in texture-like non-Gaussian models. Displayed are the predictions
from the SC model including $s_{2,3}$ (short-dashed line) and $s_{2,4}$ 
(long-dashed), which is dominated by the Gaussian contribution 
(shown as the dotted line).
The N-body simulation for 
$\sigma_8 = 1.0$ (filled triangles) shows a non-linear variance below
the linear one on all quasi-linear scales in remarkable agreement
with the SC prediction for the first non-linear contribution $s_{2,3}$.}
\label{x2text}
\end{figure}

\begin{figure} 
\centering
\centerline{\epsfysize=8.truecm 
\epsfbox{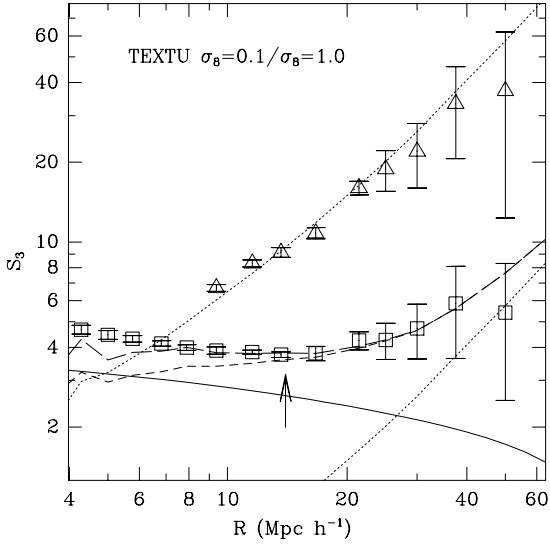}}
\caption[junk]{The hierarchical skewness, $S_3$,
for texture-like non-Gaussian models.
The triangles show the initial conditions ($\sigma_8=0.1$),
which are fitted well by the dimensional scaling $S_3= B_3/\sigma$,
shown as the upper dotted line. Squares show $S_3$
for a later output: $\sigma_8=1.0$.
The SC predictions for the $\sigma_8=1$ output are shown
as short-dashed (including the second order contribution) 
and long-dashed line (including the third order).
The continuous line shows the corresponding tree-level PT prediction for
Gaussian initial conditions. The lower dotted lines correspond to
the linear theory prediction.}
\label{s3textpt}
\end{figure}

\begin{figure} 
\centering
\centerline{\epsfysize=8.truecm 
\epsfbox{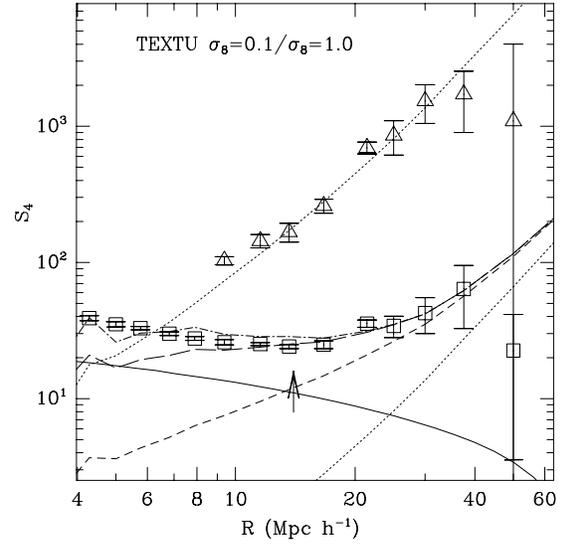}}
\caption[junk]{The hierarchical kurtosis, $S_4$,
for texture-like non-Gaussian models. Symbols and lines
as in Figure \ref{s3textpt}. The new line in the plot (dot
long-dashed) displays the SC prediction including the 4th perturbative 
contribution. In this case
the initial conditions are fitted to the corresponding
dimensional relation $S_4 = B_4 /\sigma^2$.}
\label{s4textpt}
\end{figure}

The above predictions  are in  good agreement with the N-body simulations
depicted in Figs \ref{x2text}-\ref{s4textpt}.
The N-body results (symbols with error-bars from 3 realizations) 
are from Gazta\~{n}aga and M\"{a}h\"{o}nen (1996). 
A non-linear sigma model was used for the texture dynamics, 
which was  stopped at $\sigma_8 = 0.1$. 
Density perturbations were mapped by $100^3$ particles to produce
the initial conditions for a gravitational N-body simulation.
The density fluctuations were then evolved by a $\rm P^3M$-code 
(Efstathiou \&  Eastwood 1981, Efstathiou \etal 1988)
until they reach $\sigma_8 = 1.0$. 

We first fit a model for initial conditions $\xi_J(0)$ using the
statistics of the initial density fluctuations in the simulations
obtained from a map (which assumes $\Omega=1$) 
of the texture dynamics (non-linear sigma model)
at  $\sigma_8 = 0.1$ (\eg $10$ expansion factors before 
now: $\sigma_8 = 1.0$). For the initial variance we use a numerical 
fit, which will be scaled with $a=D$ to be the linear variance. 
We corrected this initial values from the grid shot-noise, and the results 
are the average of 3 realizations, which explains why the SC predictions
are not smoothed in the Figures. For the higher-order cumulants of
the IC,
a good match for $R \simgt 10 \Mpc$ is given by the dimensional scaling
$\xibar_J = B_J \sigma^J$,
with $B_J \simeq 0.5$. This model for the initial conditions
 is shown as the upper dotted lines in the plots for the $S_J$ amplitudes
(only larger scales are shown
for clarity, smaller scales are dominated by large shot-noise
fluctuations).

For $n \simeq -1$ ($\gamma \simeq -2$), the prediction for $B_J =0.5$
is,
\bea
\sigma^2 &\approx& \sigma_l^2 - 0.02 \, \sigma_l^3 \nn \\
B^{NL}_3 &\approx& 0.5 + 3.1 \,\sigma_l  \nn \\
B^{NL}_4 &\approx& 0.5 + 6.2 \, \sigma_l + 19.7\, \sigma_l^2  
\label{bj_text}
\eea
We remark that a PT prediction 
for $B^{NL}_J$ must keep at least $J-2$ corrective terms to the linear 
prediction to include the (hierarchical) purely gravitational term, $S_J^G$
(see Eq.[\ref{dim_rat}]). This means that to yield a meaningful
prediction for the skewness and kurtosis we have to include one
and two (non-linear) corrective terms, respectively, 
to the linear theory prediction.
In practice, $n$ (or, equivalently, $\gamma$) is a function of the scale
$R$, so that the coefficients of the non-linear corrections above
(\eg $s_{2,3}$) are a function of the local slope, $\gamma$.
This slope is obtained 
numerically from the initial variance.

For the variance, $\xi_2 \equiv \sigma^2$, we find in Figure \ref{x2text}
 a good agreement between the 
the simulation output (for $\sigma_8 = 1.0$, open squares)
and the SC prediction for the leading contribution
to the non-linear variance: $s_{2,3}$ 
(short-dashed line).  The agreement
extends up to the scales where the shot-noise dominates the correlations
(beyond the point where $\sigma \approx 0.5$).
In particular, the negative contribution predicted for the variance 
in $s_{2,3}$ is clearly in agreement with the behavior 
measured in the simulations. The latter is related to
the fact that, contrary to the Gaussian case (see Paper I, \S 5.2),
the variance has a critical effective index, $n= -8/7$ for 
which the non-linear contributions change sign (see Fig \ref{scvpttop1}). 
This has nothing to do with 
the previrialization effect found for GIC (since the SC model is 
unable to account for this effect,
see Paper I, \S 5.2) but rather with the fact that the variance is still
dominated by the IC which push the variance towards lower values.
The prediction for the next contribution to the variance $s_{2,4}$ for the
NGIC (long-dashed line) and the Gaussian IC (dotted line), 
shows that this term is dominated
by the Gaussian contribution (that independent of $B_J$), 
with  a  behavior similar to the one found for GIC (see Paper I,
\S 5.2), \ie positive
contributions to the variance on all scales.
A clear departure of the variance including the first 
non-linear correction ($s_{2,3}$), with respect to the 
prediction once the second correction is included ($s_{2,4}$)
suggests the point beyond which the perturbative approach should
break down (see arrow in Fig \ref{x2text}).
The second non-linear ($s_{2,4}$) correction to the variance, seems
to be dominated by the tidal contributions missed in the SC
model prediction, as displayed in Fig \ref{x2text}.
Note nevertheless 
the similarity of this effect to the one
found in Figure 6 of Paper I. There, the first order 
(non-linear) correction to the variance in the APM simulations also 
gives a much better agreement than the
second order correction. This might indicate that 2-loop calculations
have stronger tidal contributions, so that the SC model is a poorer
approximation for higher-order loops. 

On the other hand, in the plots concerning the hierarchical amplitudes $S_J$,
the lower dotted lines are the linear theory predictions, which
only approach the non-linear results 
at the larger scales, $R\simeq 40 \Mpc$ ($\sigma \simlt 0.1$),
where they do better than the GIC predictions, but are not
as good as the NGIC predictions.

There is an excellent
agreement  between the SC model predictions for $S_J$ with those
from the N-body simulations up to the point where the 
prediction including the $S_{J,1}$ contribution
significantly deviates from that up to the hierarchical term, $S_{J,0}$  
(which includes the
purely gravitational term, $S_G^J$). This means that for the
skewness, $S_3$ (kurtosis, $S_4$), the SC result is reliable as long as the
prediction including the 2nd (3rd) and the 3rd (4th) perturbative 
contributions are in rough agreement.  
This point is reached approximately
for $\sigma \simeq 0.5$, which is shown as an arrow in Figs.
\ref{s3textpt} \& \ref{s4textpt}.
Notice that for NGIC,
loop corrections enter at the same order (in $\sigma$) 
as tree-level corrections.
Note also the deviation from the Gaussian prediction on quasi-linear scales 
which shows up as a 
shift from the Gaussian values, as mentioned above and predicted
in Eq[\ref{sj_dim}] (order zero in $\sigma$, $S_{J,0}$).

Similar results are found for higher-order moments, which have also been used
to check the initial correlations, $B_J \simeq 0.5$ up to $J=6$.

\subsection{Comparison with Exact PT}

The analytic PT calculations available in the literature for Non-Gaussian 
initial conditions refer only to the variance, $\sigma$, the
skewness $S_3$ (see FS94) and the  dimensional 
kurtosis, $B_4$ (see CB96) for the {\it unsmoothed} fields. We will now
compare our results to those of the exact ({\em non-local}) calculation.

\subsubsection{The Variance in exact PT}

For the unsmoothed variance  FS94 find:
\beq
\sigma^2\, = \sigma_l^2\,+\,{13\over 21}\,\xi_3^l
\,+\,{4\over 7}\,I[\xi_3^l]\,+\,\Or \left(\sigma_l^4 \right)
\eeq
where $\xi_3^l$ is the linear 3-point function: $\xi_3^l \equiv 
\langle \delta_l^3 \rangle_c = D^3 \xi_3(0)$ and
$I[\xi_3^l]$ stands for some 
multipole integral of the linear
three-point function contributing to the one-loop order. 
In particular, we define,
\beq
I[f(\x, \x \prime)] \,=\,\int {d^3 {\x} \over 4 \pi}\,
{d^3 {\x \prime} \over 4 \pi}\,{(3 \cos^2 \theta-1) \over \x^3 {\x \prime}^3}\,
f(\x,\x \prime) ,
\eeq
with $\cos \theta = \x {\x \prime}/x {x \prime}$.
That integral term,  $I[\xi_3^l]$,
competes, at the same order, with
the tree-level amplitude in the perturbative expansion for dimensional IC.
As a result, though the 
tree-level amplitudes remain (local) shearless, the tidal one-loop contribution 
cannot be exactly recovered
by the SC model. In particular, from the SC model we found for the evolved
variance, 
\beq
\sigma^2\, = \sigma_l^2\,+\,\left({S_3^G\over 3} -1\right)\,B_3\,
\sigma_l^3\,+\,\Or \left(\sigma_l^4 \right) ,
\label{varsc}
\eeq 
where here $B_J = \xi_J^l/\sigma_l^J$ denote the
dimensional amplitudes, and $S_3^G = 34/7$ (the tree-level skewness
for GIC). Replacing the latter 
number in the expressions above,
we see that our results within the SC model are able to recover
all the terms given by the exact calculation up to the one-loop 
tidal-induced term dependent on the initial three-point function, 
\eg ${4\over 7}\,I[\xi_3^l]$. 
Note that SF94 typically find (see their Table 1) that the coefficients
$\eta_J \equiv I[\xi_J]/\xi_J$ resulting from the multipole integration
are smaller than unity, $\eta_J \simlt 0.2$,
 thus typically smaller than the monopole
contribution . As mentioned above this is not totally surprising 
as correlations typically decrease with distance,
thus reducing the contribution in non-spherical geometries.
Furthermore, note that the SC results above provides with the
smoothed generalization of the FS94 results (up to tidal effects),
\eg:
\beq
\langle \delta^2 \rangle = \sigma_l^2\,+\,\left({13\over 21}
+{\gamma\over 3}\right)\, \xi_3^l \,+\, \cdots
\eeq

The general behavior found for the variance 
seems to be in qualitative agreement with the conclusions drawn 
by FS94  which pointed out 
that a larger (lower) variance than the Gaussian one is expected for models
with positive (negative) initial skewness (see also Moscardini \etal 1993).
But note that this trend changes when $S_3^G<3$, i.e., when $s_{2,3}<0$, 
which is
the leading term in the variance (see Eq.[\ref{sj_dim}]). This was not detected
by FS94 because they did not include smoothing effects.
Thus, for $n > -8/7 \approx -1.14$ or $\gamma < -1.85$, 
there is a change in this
trend, with lower variance for more positive skewness. In observations
and also in CDM models (see Gazta\~naga \& Baugh 1998), 
the spectral index is $n \simlt -1$ in
the weakly non-linear scales, going to $n> -1$ on large scales.
This will produce a characteristic change in the
shape (see Figure \ref{x2text}).

Within SC model the reason for this qualitative change
(lower or higher variance for more positive skewness) lies in the
combination of smoothing effects and the
transformation from Lagrangian to Eulerian coordinates, which reduce
non-linearities. Indeed, the negative term in
$s_{2,3} = ({S_3^G}/3-1)\,B_3$, which makes the cancellation
possible, comes from the  Lagrangian to Eulerian transformation. 
While
smoothing effects, can reduce the value of $S_3^G$ and change the
relative sign.  

Note the repeated appearance of a
a  critical index  $n \simeq -8/7 \approx -1.14$  
($\gamma \simeq -1.85$) in the SC model predictions for the cumulants
(see also Paper I).
At this index, the corrections to the variance are minimal for GIC
and change sign for NGIC. Furthermore, the skewness coefficient $S_{3,0}$
also shows a critical behavior (see \S 3.2) for that particular value.
This critical index happens to correspond to the solution to the equation 
$S_3^G=3$, 
which lies close to the observational values of both
$S_3$ (see Gazta\~naga 1994) and $n$ itself
(Gazta\~naga \&  Baugh 1998) in the quasi-linear regime.

\subsubsection{Skewness \& Kurtosis in exact PT}

Similarly, for the skewness, FS94 give,
\bea
S_3 &=& {{\xi_3^l}\over{\sigma_l^4}}\,+\,{34\over 7}\,+
\, {10\over 7}\,{{\xi_4^l}\over{\sigma_l^4}}\,-\,
{26\over 21}\,{(\xi_3^l)^{2}\over{\sigma_l^6}}\,
\nn \\
&+& {\,{6\over 7}\,{I[\xi_4^l]\over \sigma_l^4}}\,-\,{8\over 7}\,
{{\xi_3^l\,I[\xi_3^l]}\over{\sigma_l^6}}\,+
\,\Or \left(\sigma_l \right)
\label{eq:exacts3}
\eea
with the tidal terms expressed as 
integrals $\sim I[\xi_3^l],
I[\xi_4^l]$ of the linear three- and four-point functions respectively.
Recalling the corresponding expression for the skewness in the SC model, 
\beq
S_3 = {B_3\over \sigma_l} +S_3^G - 2\,\left({S_3^G\over 3}-1\right)\,B_3^2 +  
\left({S_3^G\over 2}\,-1\right)\,B_4\,+ \dots 
\label{s3sc}
\eeq
we see again how the SC approximation exactly recovers all the local terms,
but not the tidal or non-local integrals.
Note that, in this case, 
the two  tidal contributions to the reduced cumulants appear at the one-loop order
as differences in the $J$-point functions, leading in most of the cases (see
FS94, Table 1) to a marginal contribution to the SC prediction. 

 Finally, we turn to the result derived by CB96,
for the dimensional (non-linear) kurtosis, $B_4^{NL}$ for an arbitrary 
Non-Gaussian density field. They find,
\beq
B_4^{NL} = B_4\,+\,B_{4,1}\,\sigma_l\,+\,\Or \left(\sigma_l^2 \right) 
\eeq
being,
\bea
B_{4,1} &=& {47 \over 21}\,B_5\,+\,{136 \over 7}\,B_3\,-\,{26\over 21}\,B_3\,B_4\,
\nn \\
&+& {8\over 7}\,\left(B_5\,{{I[\xi_5^l]}\over{\sigma_l^5}}\,-\,B_4\,B_3\,
{{I[\xi_3^l]}\over{\sigma_l^3}} \right)
\label{bj_pt}
\eea


From the SC  approximation we find a similar expression 
(see Eq.[\ref{dim_rat}]), 
which after replacing $S_3^G = 34/7$, 
matches the PT result for the tree-level contribution 
while failing to
reproduce the one-loop tidal 
terms. These involve some
integrals of the initial $J$-point functions, just as the case of
the skewness commented above. Again, we expect this tidal contribution
to generically yield a marginal net correction to the SC model prediction
($\eta_J \simlt 0.2$).


Note again that all the exact results mentioned above correspond only to the
{\em unsmoothed} cumulants. In the present context, the local contribution
can be easily extended for the smoothed case.
The {\em smoothed} predictions are implicitly given in
equations (\ref{sj_dim}) and Table \ref{ngicsc} and
tend to yield smaller non-linearities, similar to what was found in Paper I
for GIC. 

\subsection{Predictions for the isocurvature CDM cosmogony}
\label{sec:iso}

As an interesting working example, 
consider the isocurvature CDM cosmogony presented
recently by Peebles (1998). In the particular model presented
Peebles used for the initial conditions the following parameters:
$B_3 =D_3 \simeq 2.5$, $B_4 = D_4 \simeq 9.9$ and 
$\gamma= -2 \epsilon = -1.2$.
From Eqs.[\ref{sj_dim}],[\ref{dim_rat}] we have that 
the leading order non-linear corrections to $\sigma^2$, $B^{NL}_3=S_3 \sigma$ 
and $B^{NL}_4=S_4\sigma^2$, in terms of
the linear {\em rms} fluctuation, $\sigma_l$, are:
\bea
\sigma^2 &\approx& \sigma_l^2 + 0.55 \, \sigma_l^3 \nn \\
B^{NL}_3 &\approx& 2.5+ 9.8 \,\sigma_l  \nn \\
B^{NL}_4 &\approx& 9.9 + 98 \, \sigma_l + 560 \, \sigma_l^2   
\label{bj_icdm}
\eea
where to estimate the non-linear correction for $B^{NL}_4$  
we have assumed $B_5 \simeq 1.6 (B_4)^{3/2} \simeq 50$,
and  $B_6 \simeq 3.3 (B_4)^2 \simeq 300$,
following the $\chi^2$ distribution.
These non-linear corrections are not very sensitive to the values
assumed for $B_5$ and $B_6$, as setting them to zero (a very conservative
assumption) yields
$B^{NL}_4 \approx 9.9 + 26 \, \sigma_l + 250 \, \sigma_l^2$,
which is anyway a very large correction to the linear theory
prediction, $B_{4,0} = 9.9$, when $\sigma_l \simeq 1$. 
We stress that a PT prediction 
for $B^{NL}_J$ must keep at least $J-2$ corrective terms to the linear 
prediction to include the non-negligible purely gravitational term, $S_J^G$
(see Eq.[\ref{dim_rat}]).
This can also be seen in Figs \ref{s3textpt} \& \ref{s4textpt} where the
SC prediction follows the N-body predictions up to $\sigma_l \simlt 1$
only when the first and second corrective terms are included in the
perturbative expansion of the skewness and kurtosis, respectively.

Note that the Isocurvature CDM model has a much larger 
non-linear correction for the variance than that of the
texture model discussed in \S\ref{ng_nbody} 
(see Eq.[\ref{bj_text}] and also Fig \ref{x2text}), 
because there, the spectral index, 
is lower ($n \approx -1$) than in the Isocurvature CDM cosmogony 
($n = 2 \epsilon -3 \approx -1.8$, see Peebles 1998).
According to the above results, 
the non-linear corrections to $B^{NL}_J$ are 
very large (and positive) and linear theory is no longer a good approximation
for $\sigma_l \simeq 1$. This is very similar to the situation
presented in Figures \ref{s3textpt}-\ref{s4textpt} for the texture model, 
where linear theory only works 
for $R > 50 \Mpc$, or $\sigma_l < 0.1$. 
What is more, the non-linear corrections to the $B^{NL}_J$ are typically an
order of magnitude larger than the linear values 
(see Eqs.[\ref{bj_text}],[\ref{bj_icdm}]).

These non-linear corrections
should be taken into account when comparing the model with 
observations in the galaxy catalogues, as in Table 1 of Peebles (1998),
where even at the larger scales  $\theta \simeq 1$ deg ($R \simeq 10 \Mpc$), 
one has $\sigma_l \simeq 1$. Thus, for the parameters shown in 
Eq.[\ref{bj_icdm}],
the model seems to be incompatible with current observational constraints
from galaxy catalogues.

\section{Discussion and Conclusions}
\label{discuss}

For Gaussian initial conditions (GIC), we 
found in Paper I, that the Spherical Collapse (SC) model 
gives the exact tree-level contribution to PT. It was shown that 
this contribution can be derived 
by means of a local-transformation of the IC, what is much simpler than
the vertex generating function formalism developed by 
Bernardeau (1994a, 1994b). It was also seen in Paper I that 
the SC model (in Lagrangian space) also gives
an excellent agreement for the hierarchical amplitudes $S_J$
in the loop corrections, as compared
with the results derived by Scoccimarro \& Frieman (1996a, 1996b)
for the exact PT in the diagrammatic approach. 

We stress the importance of applying the SC 
approximation in Lagrangian space. 
There, the SC model is described by a transformation
that {\em only} depends on the value of the linear field at the same point
(what we call a {\em local-density transformation}). However, when
going back to Euler space the density fluctuation (defined at a point) 
is normalized with a factor which is a function of the (non-linear)
variance. Since the variance is
a volume average of the two-point correlation function, this factor 
yields some {\em non-local} contribution to the cumulants 
(in Euler space).
This {\em non-local} contribution
is missed when introducing the SC model in Euler space {\em directly},
thus is not surprising that the predictions for the 
cumulants in the SC approximation in Euler space are a poor estimation of those
in exact PT, as the latter are dominated by the
non-local (tidal) effects (see Table A2 in Paper I). 
	
For the predictions within the SC model (in Lagrangian space) 
for the hierarchical ratios, $S_J$,
tidal effects partially cancel out as they seem to contribute
roughly hierarchically to the cumulants (see Appendix A2 in Paper I). 
This is also true for the SC model in
Euler space, but the dominance of non-local effects in the cumulants there,
yield significantly different $S_J$ ratios to those in exact PT 
(see Appendix A2 in Paper I).
Smoothing effects do not alter
substantially this interpretation (at least for a top-hat window).

As the SC model can be
easily extended to the smoothed fields, 
we were able compare the predictions for the higher-order
moments from the SC model to those measured in 
CDM and APM-like N-body simulations,
and they turned out to be in  
very good agreement in all cases up
to the scales where $\sigma_l \approx 1$, supporting our view that
the tidal effects only have a marginal contribution to the reduced
cumulants. Furthermore, the
break down of the shearless approximation roughly coincides with the regime
for which the perturbative approach itself 
breaks down, $\sigma \simeq 0.5$. That is,
where the contribution of the second and the third perturbative 
order in the SC model are significantly different.

For non-Gaussian initial conditions (NGIC) the SC
recovers all tree-graphs exactly, including all contributions
given by the exact PT 
for the variance, $\sigma$, skewness, $S_3$, and kurtosis, $S_4$,
 up to some non-local integrals, $I[\xi_J]$, involving
$J$-point initial functions. These last integrals arise as a result 
of the coupling between the
asymmetric initial conditions with the tidal forces. We argue that for 
the hierarchical ratios, $S_J$, these non-local terms 
are sub-dominant and tend to compensate each other (\eg
$I[\xi_J] < \xi_J$).

The measured higher-order moments in the N-body simulations 
with NGIC (with dimensional, texture-like, scaling)
turned out to be in good agreement with predictions from the
SC model up to the scales for which the perturbative series breaks down
(\eg see Figures \ref{x2text}-\ref{s4textpt}). As mentioned in Paper I
there is  
a critical index, $n_{\star} \,\in [-1,-2]$, for which tidal effects 
vanish and the SC is a good approximation even for the variance. 
This might explain why
we find such  a good agreement for the variance as compared to
the simulations (which have $n \simeq -1$ on weakly non-linear scales).
This good agreement is found when
using only the leading-order correction
$\Or[\sigma_l^3]$, and breaks down after including the next term,
$\Or[\sigma_l^4]$ (see Figure \ref{x2text}). This effect is similar 
to what we found for GIC when comparing the variance between SC 
predictions and simulations (Figure 6 of Paper I). This could
indicate that 2-loop calculations
have stronger tidal contributions, so that the SC 
is a poorer approximation for higher-order loops.

The general behavior found for the variance 
seems to be in qualitative agreement with the conclusions drawn 
by FS94  which pointed out 
that a larger (lower) variance than the Gaussian one is expected for models
with positive (negative) initial skewness (see also Moscardini \etal 1993).
But note that this trend changes when $S_3^G<3$ 
i.e., when $s_{2,3}<0$, 
which is
the leading term in the variance (see Eq.[\ref{sj_dim}]). This was not detected
by FS94 because they did not include smoothing effects.
Thus, for $n > -8/7 \approx -1.14$ or $\gamma < -1.85$, 
there is a change in this
trend, with lower variance for more positive skewness (see \S 3.4). 

The SC model predictions for $S_J$ show  
how the NGIC evolve slowly towards the 
(Gaussian) gravitational predictions but, even at $\sigma_8=1$,
are still significantly larger. They
show a characteristic minimum with a sharp increase in
$S_J$ with increasing scales, just as found by
Gazta\~{n}aga \& M\"{a}h\"{o}nen (1996) in N-body simulations of NGIC.
The latter agreement strongly constrains the tidal contributions 
to the higher-order
moments and gives further support to the domain of applicability of the
SC model in PT.
Thus, the resulting SC predictions for the NGIC
show a non-trivial time evolution that can be used to 
strongly discriminate
models of structure formation (see \S\ref{sec:iso}).

\section*{Acknowledgments}

We want to thank Roman Scoccimarro, Francis Bernardeau
and Josh Frieman 
for carefully reading the 
manuscript and pointing out useful remarks. 
EG acknowledges support from CIRIT (Generalitat de
Catalunya) grant 1996BEAI300192. 
PF acknowledges a PhD grant supported by 
CSIC, DGICYT (Spain), projects PB93-0035 and PB96-0925.
This work has been
supported by CSIC, DGICYT (Spain), projects
PB93-0035, PB96-0925, and CIRIT, grant GR94-8001.

\section{References}

\def\refe {\par \hangindent=.7cm \hangafter=1 \noindent}
\def\aj { ApJ, }
\def\aa {A \& A, }
\def\prl {Phys.Rev.Lett., }
\def\ajs{ ApJS, }
\def\mn { MNRAS, }
\def\apl { ApJ.Lett., }

\refe Bernardeau, F., 1992, \aj 392, 1 
\refe Bernardeau, F., 1994a, A\&A 291, 697 
\refe Bernardeau, F., 1994b, \aj 433, 1 
\refe Chodorowski, M.J., Bouchet, F.R., 1996, \mn, 279, 557 (CB96) 
\refe Efstathiou, G., Eastwood, J.W., 1981,  \mn 194, 503
\refe Efstathiou, G., Frenk, C.S., White, S.D.M., Davis, M. 1988, \mn 235, 715
\refe Fosalba, P., Gazta\~{n}aga, E., 1998, \mn in Press 
(Paper I, this issue). 
\refe Fry, J.N., 1984, \aj 279, 499
\refe Fry, J.N., Gazta\~naga, E., 1993, \aj 413, 447 
\refe Fry, J.N., Scherrer, 1994, R.J., \aj 429, 1 (FS94)
\refe Gazta\~naga, E. \&  Baugh, C.M., 1998, \mn 294, 229
\refe Gazta\~naga, E. \& M\"{a}h\"{o}nen, P., 1996, \apl 462, L1
\refe Goroff, M.H., Grinstein, B., Rey, S.J., Wise, M.B., 1986, \aj 311, 6
\refe Jain, B., Bertschinger, E., 1994, \aj 431, 495
\refe Juszkiewicz, R., Bouchet, F.R., Colombi, S., 1993 \apl 412, L9
\refe Moscardini, L., Coles, P., Matarrese, S., Lucchin, F., 
Messina, A., 1993, \mn  264, 749
\refe Peebles, P.J.E., 1980, {\it The Large Scale Structure of the 
Universe:} Princeton University Press, Princeton
\refe Peebles, P.J.E., 1998, \aj submitted, astro-ph/9805212
\refe Silk J., Juszkiewicz, R.  1991, Nature, 353, 386 
\refe Scoccimarro, R., Frieman J., 1996a, \ajs 105, 37 
\refe Scoccimarro, R., Frieman J., 1996b, \aj 473, 620 
\refe Turok, N., Spergel, D.N., 1991, \prl 66, 3093 
\refe Weinberg, D.H., Cole, S., 1992, \mn 259, 652

\end{document}